\documentclass{article}

\usepackage{subfigure}
\usepackage{makecell}
\usepackage{stfloats}
\usepackage{PRIMEarxiv}

\usepackage[utf8]{inputenc} 
\usepackage[T1]{fontenc}    
\usepackage{hyperref}       
\usepackage{url}            
\usepackage{booktabs}       
\usepackage{amsfonts}       
\usepackage{nicefrac}       
\usepackage{microtype}      
\usepackage{lipsum}
\usepackage{fancyhdr}       
\usepackage{graphicx}       
\graphicspath{{media/}}     
\usepackage{amsmath}
\title{The Impact of 2D and 3D Gamified VR on Learning American Sign Language
}

\author{
  Jindi Wang  \\
  Durham University \\
  \texttt{jindi.wang@durham.ac.uk} \\
  \And
  Ioannis Ivrissimtzis \\
  Durham University  \\
  \texttt{ioannis.ivrissimtzis@durham.ac.uk} \\
   \And
  Zhaoxing Li \\
  University of Southampton \\
  \texttt{zhaoxing.li@soton.ac.uk} \\
   \And
  Lei Shi \\
  Newcastle University \\
  \texttt{lei.shi@newcastle.ac.uk} \\
}

\begin{document}
\maketitle

\begin{figure}[ht]
    \centering
  \includegraphics[width=0.51\linewidth]{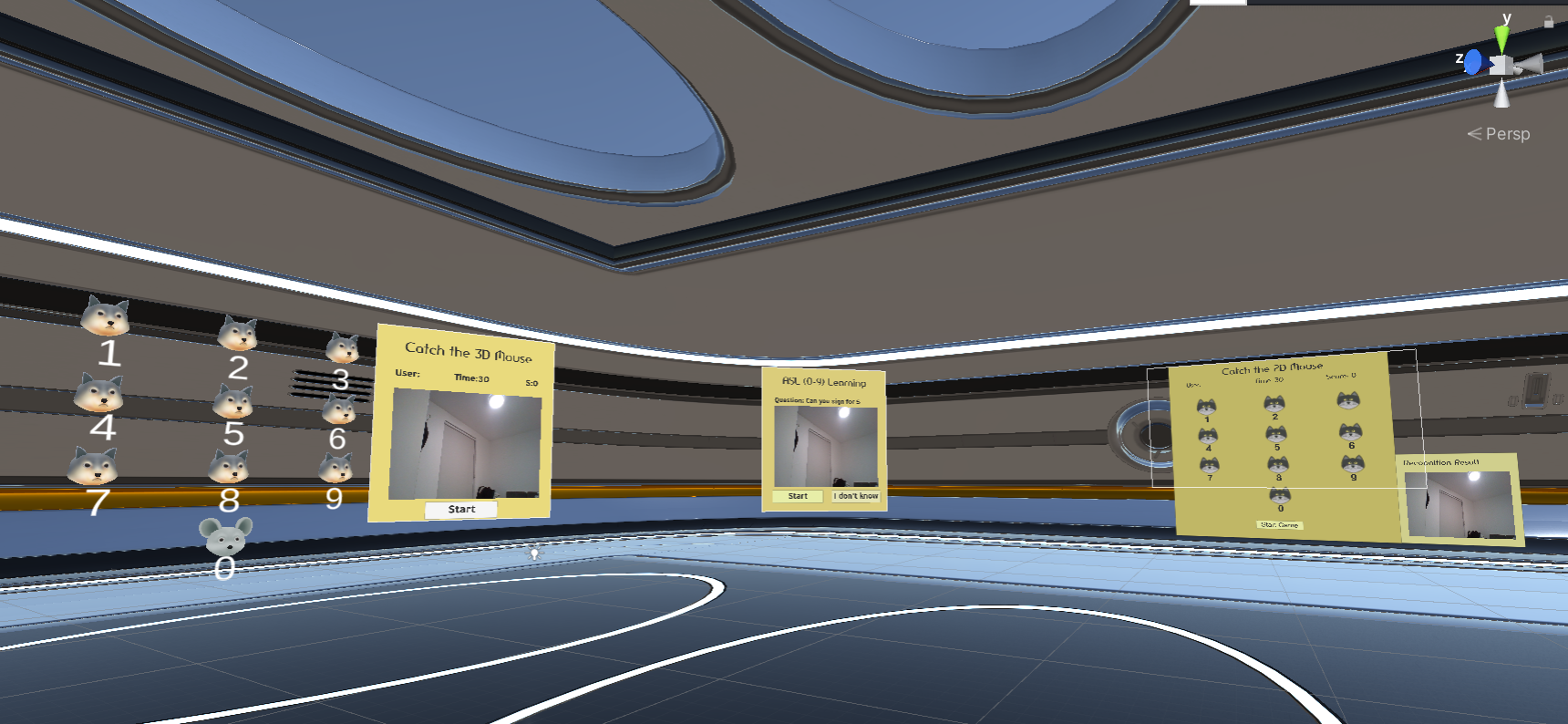}
  \includegraphics[width=0.485\linewidth]{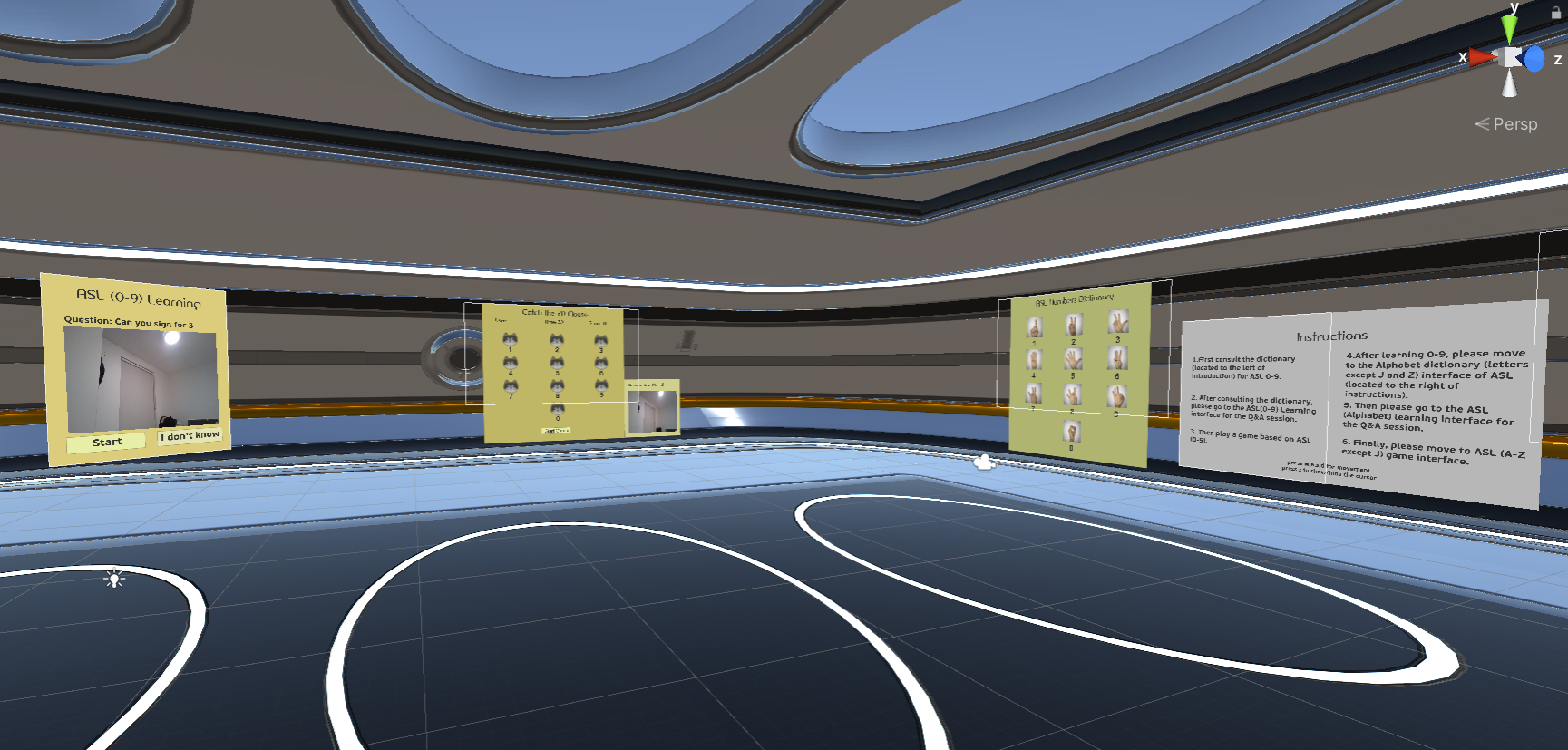}
  \caption{The implemented 2D versus 3D gamified virtual environment for ASL learning.}
  \label{fig:environment}
\end{figure}
  
\begin{abstract}
Sign language has been extensively studied as a means of facilitating effective communication between hearing individuals and the deaf community. With the continuous advancements in virtual reality (VR) and gamification technologies, an increasing number of studies have begun to explore the application of these emerging technologies in sign language learning. This paper describes a user study that compares the impact of 2D and 3D games on the user experience in ASL learning. Empirical evidence gathered through questionnaires supports the positive impact of 3D game environments on user engagement and overall experience, particularly in relation to attractiveness, usability, and efficiency. Moreover, initial findings demonstrate a similar behaviour of 2D and 3D games in terms of enhancing user experience. Finally, the study identifies areas where improvements can be made to enhance the dependability and clarity of 3D game environments. These findings contribute to the understanding of how game-based approaches, and specifically the utilisation of 3D environments, can positively influence the learning experience of ASL.
\end{abstract}


\section{Introduction}\label{sec:Introduction}

Learning American Sign Language (ASL) has significant advantages that go beyond those experienced by the deaf community. First and foremost, knowing ASL makes it easier to communicate effectively with people who are deaf or hard of hearing, promoting inclusion and lowering obstacles. The pursuit of ASL proficiency shows a dedication to building an inclusive society that values diversity and guarantees equal access for all. Learning ASL also offers unique linguistic and cognitive benefits \cite{newport2017acquisition, wang2023user}. According to a large body of research, learning sign language improves cognitive functions \cite{freel2011deaf,li2023towards,li2021survey} and heightens visual-spatial awareness \cite{quinto2017case,li2023sim}. Being proficient in ASL also provides access to prospective job opportunities given the rising demand for ASL interpreters, educators, and other professionals who can bridge the communication gap between the deaf and hearing communities. As a result, learning ASL has become a significant area for educational study.

Virtual reality (VR) and gamification technologies are rapidly advancing, and their potential for revolutionising sign language learning is gaining significant attention from researchers. The integration of gamified approaches in sign language education offers a range of compelling advantages, including enhanced interactivity, immersive practice experiences, and heightened learning motivation. One notable advantage of gamified sign language learning is the provision of interactive practice opportunities. Traditional sign language learning often relies on static materials, such as textbooks or videos, which limit learners' ability to actively engage with the language. By incorporating gamification elements, learners can participate in dynamic and interactive exercises, allowing them to practice their signing skills in a simulated environment. This interactivity fosters a more engaging and hands-on learning experience, resulting in improved retention and fluency \cite{oluwajana2019adoption, buckley2016gamification, wang2023exploring, roosta2016personalization}. Additionally, gamified sign language instruction offers a plethora of engaging and practical opportunities that allow students to put their sign language understanding and proficiency to use in a relaxed and pleasant environment. Students can enhance their sign language communication abilities by interacting with the game's characters and completing sign language puzzles, and they can keep becoming better with practice. Students' engagement and learning effectiveness are both improved by this interactive and practical teaching method \cite{shohieb2019gamified, wang2024comparative, wang2024impact, antonaci2019effects,shi14Con,shi16mo}.

As a result, extensive research has been conducted to gamify the process of learning sign language, primarily emphasising the creation and advancement of 2D and 3D games. For example, CopyCat, a 2D game proposed by Zafrulla \textit{et al.} \cite{zafrulla2011copycat} is specifically designed to engage young children in interactive computer-based ASL learning, employing cutting-edge gesture recognition techniques. Its user-friendly interface comprises tutorial videos demonstrating accurate signs, a live video enabling gesture input for the recognition system, and real-time feedback on the child's progress. Additionally, the game features a character named Iris the cat, who carries out the child's instructions. Moreover, an educational game called MatLIBRAS Racing \cite{pontes2020educational}, is designed to teach sign language for natural numbers from a cognitive perspective. The game received positive feedback from students in terms of its educational and gaming features, fostering social relationships among players and facilitating sign language learning. It suggests that MatLIBRAS Racing holds promise as an effective educational tool, particularly in academic settings. Besides, 3D games such as Economou \textit{et al.} \cite{economou2022work} presented a work-in-progress study that focuses on evaluating the impact of combining scaffolded instruction with gamification to design a 3D interactive game to support learning the sign language alphabet. Moreover, Adamo \textit{et al.} \cite{adamo2007smile} presented the second iteration of the SMILE project, which introduces an immersive 3D game tailored for both deaf and hearing children. The game's enhanced design and user interaction have been meticulously crafted to elevate motivation and appeal, fostering an engaging and captivating experience. Although the evaluations demonstrated that the application was enjoyable and user-friendly, the study did not assess the actual learning outcomes achieved by the participants \cite{wangchi2024, wang2023developing}. Collectively, the current proposals for gamified sign language learning have made valuable contributions to the field, advancing our understanding and paving the way for further developments \cite{li2024integrating, soton489244}. However, the predominant focus of these approaches has been on either 2D or 3D game formats, overlooking the exploration of potential disparities in user experiences between these two genres.

For the purposes of our study, we leveraged widely used VR technology to create a learning environment that provides users with an immersive environment to learn the ASL numbers 0-9. To improve the user experience, we developed and introduced a whack-a-mole game inspired by the sign language game ASL Sea Battle \cite{bragg2021asl,li2023deep,li2023broader} proposed by Bragg\textit{et al.}, a game that makes it easy to collect user data. In our system design, we built two versions of whack-a-mole games, called 2D whack-a-mole and 3D whack-a-mole.

To the best of our knowledge, no prior user studies have specifically examined and compared the user experience of learning ASL through 2D games versus 3D games. Consequently, we conducted a user study using a survey methodology proposed by Schrepp \textit{et al.} \cite{schrepp2014applying}, aiming to answer the research question \textbf{\textit{``What is the impact of game type (2D vs 3D) on system usability, user experience, and learning outcome in ASL learning?''}}. Our main contributions are as follows:
\begin{itemize}
    \item We have developed an immersive virtual environment that facilitates learning ASL numbers 0–9, featuring two variations of a Whack-a-Mole game. Our system offers a distinct and captivating approach to ASL learning, which we expect will enhance user satisfaction and engagement.
    \item A user study comparing the potential of 2D and 3D games in enhancing user experience and improving learning outcomes. Empirical suggests a positive impact of 3D game environments on user engagement and experience, specifically in terms of attractiveness, usability, and efficiency. Additionally, we highlight areas for improvement to ensure the dependability and clarity of 3D game environments.
\end{itemize}

\section{Related Work}\label{sec:Related Work}

\subsection{Sign Language Recognition}
Researchers have conducted several studies on sign language recognition based on deep learning and computer vision techniques. Camgoz \textit{et al.} \cite{camgoz2020sign} introduced an innovative transformer-based architecture that simultaneously learns Continuous Sign Language Recognition and Translation, eliminating the need for ground-truth timing information. This joint approach addresses two interdependent sequence-to-sequence learning problems and yields notable performance improvements. Bragg \textit{et al.} \cite{bragg2019sign} discuss the challenges and opportunities in developing effective sign language recognition, generation, and translation systems, emphasizing the multidisciplinary expertise required. They presented the comprehensive outcomes of an interdisciplinary workshop, covering essential background information, an overview of the current state-of-the-art, key challenges, and a compelling call to action for the broader research community. Pu \textit{et al.} \cite{pu2019iterative} proposed an alignment network with iterative optimisation for weakly supervised continuous sign language recognition. Their framework comprises a 3D convolutional residual network for feature learning and an encoder-decoder network with connectionist temporal classification for sequence modelling. Zhang \textit{et al.} \cite{zhang2020mediapipe} presented MediaPipe Hands, an on-device hand-tracking pipeline designed for real-time usage in augmented reality and virtual reality applications. Wadhawan \textit{et al.} \cite{wadhawan2020deep} proposed a deep learning-based system for recognising static signs in sign language using convolutional neural networks, achieving high training accuracy and surpassing earlier works that focused on a limited number of hand signs. Jiang \textit{et al.} \cite{jiang2021skeleton} proposed a Skeleton Aware Multi-modal SLR framework (SAM-SLR) that leverages multi-modal information to achieve higher recognition rates. Kumar \textit{et al.} \cite{kumar20223d} proposed a method for recognising 3D sign language using spatiotemporal graph kernels, which is signer invariant, motion invariant, and faster compared to existing graph kernel approaches.

After a thorough review of the available research on sign language recognition, we concluded that Mediapipe aligns best with the objectives of our study. Consequently, we employed Mediapipe as our tool for sign language recognition, capitalising on its exceptional accuracy in identifying hand landmark points in real-time. Moreover, as an open-source hand gesture detection framework developed by Google, Mediapipe benefits from robust support and comprehensive documentation, adding to its appeal and reliability.

\subsection{Sign Language Dictionary}
Several articles delve into various research studies focusing on sign language dictionaries. Schnepp \textit{et al.} \cite{schnepp2020human} created an animated sign language dictionary to facilitate communication between caregivers and residents who use sign language. Alonzo \textit{et al.} \cite{alonzo2019effect} conducted a study to examine the relationship between the performance of automatic sign recognition technology and users' subjective judgments when searching for unfamiliar words in ASL dictionaries. Their findings revealed that metrics incorporating the precision of the overall results list and the similarity of other words within the list demonstrated a stronger correlation with users' judgments compared to the metrics typically reported in previous research on ASL dictionary. Hassan \textit{et al.} \cite{hassan2021effect} investigated the impact of a webcam-based ASL dictionary search system on user judgments. Their study found that various factors such as the position of the desired word in the results list, whether the word appeared above or below the fold, and the similarity of other words in the list, significantly influenced users' perceptions of the system. In another study, Hassan \textit{et al.} \cite{hassan2022design} discussed the design and evaluation of a hybrid search approach for American Sign Language to English dictionaries. This approach synergistically integrated video-based queries and linguistic properties filtering, resulting in heightened search speed and accuracy. Through interviews with ASL learners and a between-subjects experiment, the hybrid search system exhibited superior performance metrics and user satisfaction compared to a video-based search system. These studies demonstrate the potential of incorporating sign language dictionary queries to enhance learning efficiency and improve the overall learning experience for users. Drawing from these findings, our learning environment design incorporated an ASL dictionary user interface, empowering users to efficiently search for sign language representations.

\subsection{Sign Language Games}
Additionally, there have been studies that specifically focus on the gamification of sign language learning. For example, Gameiro \textit{et al.} \cite{gameiro2014kinect} proposed Kinect-Sign, a serious game designed to teach non-deaf individuals sign language. The game offers two modes: School-mode for learning letter signs, and Competition-mode for testing the acquired skills, through engaging challenges. Besides, Uluer \textit{et al.} \cite{uluer2015new} presented a robotic platform designed for sign language tutoring for children with communication impairments. The platform utilises an interactive five-fingered robot called Robovie R3, which expresses selected words in Turkish sign language using hand and body movements combined with facial expressions. Moreover, Bantupalli \textit{et al.} \cite{bantupalli2018american} developed a vision-based system that translates sign language into text, aiming to enhance communication between signers and non-signers. Samonte \cite{samonte2020assistive} developed an e-tutor system to support instructors in teaching sign language. Economou \textit{et al.} \cite{economou2020using} designed a serious game to assist adults in learning sign language and bridge the communication gap between hearing-impaired and able-hearing individuals. These studies suggest that integrating gamified components into sign language instruction can enhance learners' motivation to learn. Building upon these insights, we incorporated sign language games into the user interface of our learning environment to improve users' ASL learning experiences.

Hence, our study aimed to investigate the distinct impacts of 2D and 3D games on ASL learning, providing valuable insights for future research in this domain. To achieve this, we created a VR environment utilising advanced VR technology as a platform for users to engage in ASL learning. Within this immersive learning environment, we incorporated various features such as sign language dictionaries, interactive question-and-answer sessions, and captivating sign language games to enhance the overall user experience. Among the sign language games we developed, one notable example was a whack-a-mole game, offered in both a 2D version and a 3D version. By comparing the experiences of two distinct user groups, each playing their perspective game versions, we aimed to explore the differential effects of 2D and 3D games on the process of ASL learning.

\section{Learning Environment and Games}\label{sec:System Details}
The immersive learning environment for learning ASL numerals from 0 to 9 is shown in Fig.~\ref{fig:environment}. Unity (version 2020.3.32f1) was used to create the scene. Utilising the eye-tracking functionality of the HTC Vive Pro, users interact with the device by tracking their eye location. When a user's attention is fixed for three seconds, the system allows them to click or choose an object.

For image acquisition, a built-in camera was utilised, connected to a PC that employed openCV (version 3.4.2) \cite{bradski2000opencv}. Gesture recognition was implemented using Mediapipe \cite{zhang2020mediapipe}, which detected the user's hand and extracted a sequence of 21 feature points ($p_0$, $p_1$, $p_2$,..., $p_{20}$) representing landmarks on the hand. The coordinate frame's origin was set at $p_0$, the point near the user's wrist at the bottom of the palm. The classifier employed was a multilayer perceptron consisting of three fully connected layers, implemented in Python 3.6 \cite{python2021python} and Tensorflow 2.6.0 \cite{dillon2017tensorflow}, yielding recognition accuracy rates exceeding 90\%. This level of accuracy was considered satisfactory for the study's objectives, ensuring a smooth user experience.

\begin{figure}[h]
    \centering
    \includegraphics[width=0.68\linewidth]{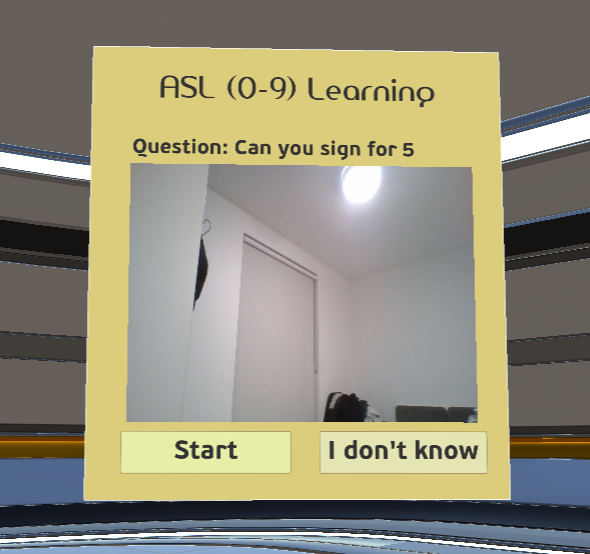}
    \caption{Quiz module for learning ASL}
    \label{fig:quiz}
\end{figure}

To enhance the effectiveness of the learning process, a question-answer module was integrated into the system. This module allows users to evaluate their proficiency level and practice their signing skills by responding to randomly generated questions from a database. In Fig.~\ref{fig:quiz}, an example is shown where the system presents the question ``Can you sign for 5?". The user has the option to refer to the dictionary or utilise their acquired skills to sign the number `5'. Alternatively, they can choose the ``I don't know" option, prompting the system to demonstrate the correct expression. In this case, the user is encouraged to continue practising until they feel confident enough to sign the digit independently by pressing the relevant button.

\subsection{2D Game and 3D Game}

\begin{figure}[ht]
    \centering
    \includegraphics[width=0.405\linewidth]{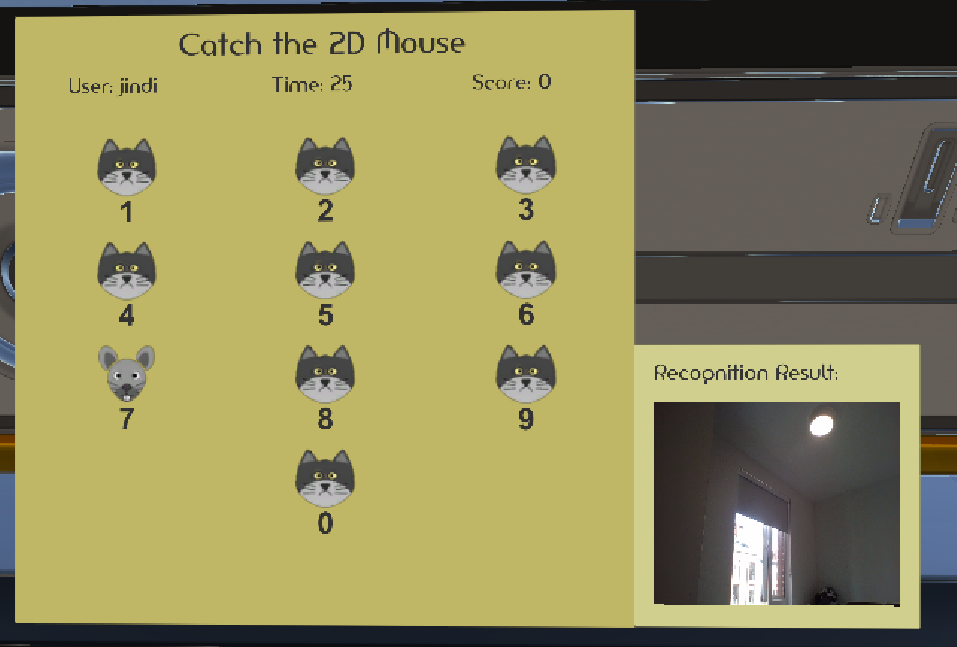}
    \includegraphics[width=0.505\linewidth]{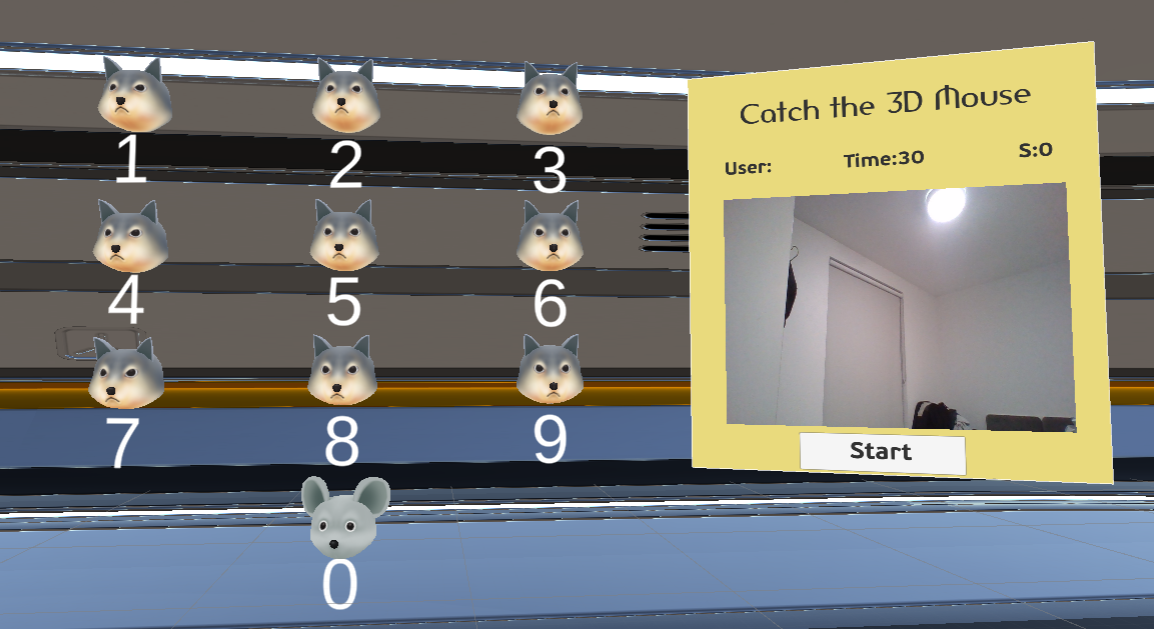}
    \caption{2D Whack-a-Mole Game (\textbf{Left}); 3D Whack-a-Mole Game(\textbf{Right})}
    \label{fig:2D and 3D games}
\end{figure}

To explore potential differences in user experiences during ASL learning, we developed both a 2D and a 3D version of our game, Whack-a-Mole, incorporating ASL elements. The objective of the game remains the same in both versions: players must promptly identify and sign the current location of the gopher within a designated time limit.

In both the 2D and 3D versions, the game interface consists of a grid-like layout where each location is uniquely identified. Players earn one point for correctly signing the position of the gopher, indicating their understanding of ASL gestures. Conversely, no points are awarded if the player fails to accurately sign the gopher's location. To maintain an engaging experience, we have implemented a time limit of 3 seconds for each round. If the player does not sign the correct location within the given time, a new gopher automatically appears, providing the player with another opportunity to score points. The game duration is set at a total of 30 seconds, creating a challenging yet manageable time frame for players to showcase their ASL recognition and signing skills.

By offering both 2D and 3D versions of the game, we aimed to investigate potential variations in user experiences and learning outcomes across the two interfaces. This allowed us to explore the potential influence of immersion and visual depth within the 3D environment on players' engagement, motivation, and overall ASL learning, compared to the 2D version. Fig.~\ref{fig:2D and 3D games} visually depicts the interfaces of both the 2D and 3D versions of the Whack-a-Mole game, showcasing the grid layout and the gopher's locations.

\section{Methodology}\label{sec:Methodology}

\subsection{Participants and Procedure}
 
The sample consisted of 24 participants, aged between 23 and 34 years (M = 28.13, SD = 2.55). All participants played the game twice and were divided into four groups, covering all possible combinations between 2D or 3D environments and first or second attempt. The aim of the design was to detect the learning effect in general and the effect of familiarisation with the virtual environment in particular. None of the participants possessed any prior experience or formal education in ASL, thus ensuring that the focus of the study was on individuals without pre-existing sign language knowledge.

In order to successfully enhance the learning process by combining instructional scaffolding approaches, we divided the learning process into three stages: \textbf{Learn}, \textbf{Practice}, and \textbf{Assess}. It is a suggested strategy for aiding ASL learning that entails the creation of a teaching tool that combines innovative technology with gamification components \cite{wood1976role}. The details of the learning process are described below:
\begin{itemize}
    \item \textbf{Learn}: Participants are encouraged to explore and navigate the virtual learning platform at their own pace, acquainting themselves with the layout and function of each user interface. Following this, they are requested to spend three minutes familiarising themselves with the 0-9 ASL expressions, utilising the 0-9 ASL dictionary interface.
    \item \textbf{Practice}: To enhance users' comprehension of 0-9 ASL, we offer a Quiz interface where users can engage in a dynamic question-and-answer session. Within the Quiz interface, participants will be presented with randomly generated questions to respond to. During this phase, participants are encouraged to actively engage in the Q\&A session for a period of three minutes.
    \item \textbf{Assess}: After the users have demonstrated their proficiency in 0-9 ASL expressions and understanding of the subject matter by successfully completing the initial two stages, we present them with an engaging game called Whack a-Mole, specifically designed to assess their learning. This stage aims to compare the effects of the 2D and 3D versions of the game on user experience. First, we asked the users to play the game once and complete our user evaluation questionnaire. In the second attempt on the game, half of the users continued on the same environment, while the other half swapped between 2D and 3D. That is, we had four groups in total, of six participants each: \textbf{2D-2D}, \textbf{2D-3D}, \textbf{3D-2D}, and \textbf{3D-3D}, depending on the environments of their first and second attempts. 
\end{itemize}

\subsection{Evaluation Methods}
Three dimensions of the VR learning environments were evaluated and compared: \textit{usability}, \textit{user experience}, and \textit{user performance}.

For \textbf{usability}, the well-established SUS (System Usability Scale) questionnaire \cite{bangor2009determining} was employed. The participants were asked to rate the 2D and 3D versions of the VR learning environments on a scale of 1 (strongly disagree) to 5 (strongly agree) after using them.

For \textbf{user experience}, we employed the user survey method proposed by Schrepp \textit{et al.} \cite{schrepp2014applying}. This approach utilises six scales, each representing a distinct aspect of the user experience: \textit{Attractiveness}, \textit{Efficiency}, \textit{Perspicuity}, \textit{Dependability}, \textit{Stimulation}, \textit{Novelty}. These scales provide a comprehensive framework for evaluating the users' perceptions and attitudes. Each scale comprises several specific items that capture various dimensions of the user experience, as outlined in Table~\ref{tab:factors}. 
To gather data on the user experience, we used a 7-point Likert scale for each of the 26 items in the questionnaire. Participants were asked to rate their level of agreement with each statement, ranging from 1 (strongly agree with a negative statement) to 7 (strongly agree with a positive statement). This rating scale allowed a quantitative analysis of the participants' perceptions across the various dimensions of the user experience.

Finally, for \textbf{user performance}, i.e., how well they learned using the VR environments, we collected and analysed the players' game scores. 
\begin{table}[ht]
  \centering
  \caption{User experience questionnaire.}
  \label{tab:factors}
  \scalebox{0.75}{
  \begin{tabular}{ |l|l| }
  \hline
  \textbf{Attractiveness} & \textbf{Perspicuity} \\
  \hline
   annoying / enjoyable & not understandable / understandable \\ 
   good / bad & easy to learn / difficult to learn \\
   unlikable / pleasing  &  complicated / easy \\
   unpleasant / pleasant & clear / confusing \\
   attractive / unattractive &  \\
   friendly / unfriendly & \\
  \hline
   \textbf{Efficiency} & \textbf{Dependability} \\
  \hline
    fast / slow &  unpredictable / predictable\\
    inefficient / efficient & obstructive / supportive\\
    impractical / practical & secure / not secure\\
    organized / cluttered & meets expectations / does not meet expectations\\
  \hline
  \textbf{Stimulation} & \textbf{Novelty} \\
  \hline
   valuable / inferior & creative / dull\\
   boring / exiting & inventive / conventional\\
   not interesting / interesting &  usual / leading edge\\
   motivating / demotivating & conservative / innovative\\
  \hline
  \end{tabular}
  }
\end{table}

\section{Results} \label{sec:Results}

\subsection{Comparing Usability}

On a scale of 0 to 100, with 0 representing low usability and 100 representing high usability, both the 2D and 3D game environments achieved high usability ratings on the SUS scale. The 2D game received an average score of 76.25 (SD = 5.15), while the 3D game scored 80.63 (SD = 3.41). These scores indicate that both game environments received ``good'' usability ratings \cite{bangor2009determining}. The low standard deviations observed for both games suggest that the majority of participants had consistent and positive experiences with the usability of the interfaces. These findings highlight the success of creating user-friendly interfaces for both 2D and 3D games, with the transition to 3D not significantly impacting usability. This suggests that both 2D and 3D games can be equally effective in terms of usability, and it is important to consider the specific needs and preferences of the target audience when designing game interfaces.

The comparable high usability scores achieved by both the 2D and 3D game environments demonstrate the success of our design in terms of intuitive controls, clear visuals, and engaging gameplay mechanics. The positive SUS ratings indicate that users found both interfaces easy to navigate and efficient. 

\subsection{Comparing User Experience}

\begin{figure}[ht]
    \centering
    \includegraphics[width=0.6\linewidth]{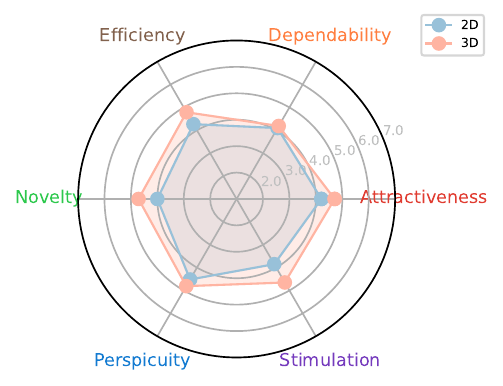}
    \caption{Six scales of user experience}
    \label{fig: six scales}
\end{figure}

Fig.~\ref{fig: six scales} presents a comprehensive comparison of user experience during ASL learning, between the two versions of the game environments. In summary, the user experience analysis reveals that both 2D and 3D games received positive ratings across various scales. However, the 3D game consistently outperformed the 2D game in several aspects, including visual attractiveness, efficiency, novelty, perspicuity, and stimulation. The users found that the visuals of the 3D game were appealing and innovative, providing a fresh and immersive gaming/learning experience, as shown in Figure 4(a) and Figure 4(d). Additionally, the 3D game demonstrated higher efficiency, offering streamlined user and interface interactions, but in terms of ``organized'', it is similar to those of the 2D game, as shown in Figure 4(c). The interface of the 3D game was also perceived as easier to learn and understand, enhancing the overall user experience, but in terms of ``easy'', it is lower than that of the 2D game, as shown in Figure 4(e), which means users found that the 2D game is easier to operate. Moreover, the 3D game succeeded in delivering a more stimulating and engaging user experience, eliciting higher levels of excitement and immersion from the users, as shown in Figure 4(f). Overall, the findings suggest that the added dimensionality and immersive elements of the 3D game contributed to a superior user experience compared to the 2D game. While the 3D game excelled in several areas, Figure 4(b) indicates that both 2D and 3D games were perceived as having low dependability by the users, as they found both games met their basic needs, but there are areas where they would like to improve, such as the complexity of the game, the difficulty of completing the level, and so on. Despite the strengths of the 3D game, it is worth noting that the 2D game still received positive ratings across all scales, indicating that it provided a satisfactory user experience as well. The results emphasise the importance of considering various factors, including visuals, efficiency, novelty, perspicuity, and stimulation, when designing game environments to enhance user experience and engagement.

\begin{figure}[ht]
    \centering
    \includegraphics[width=0.8\textwidth]{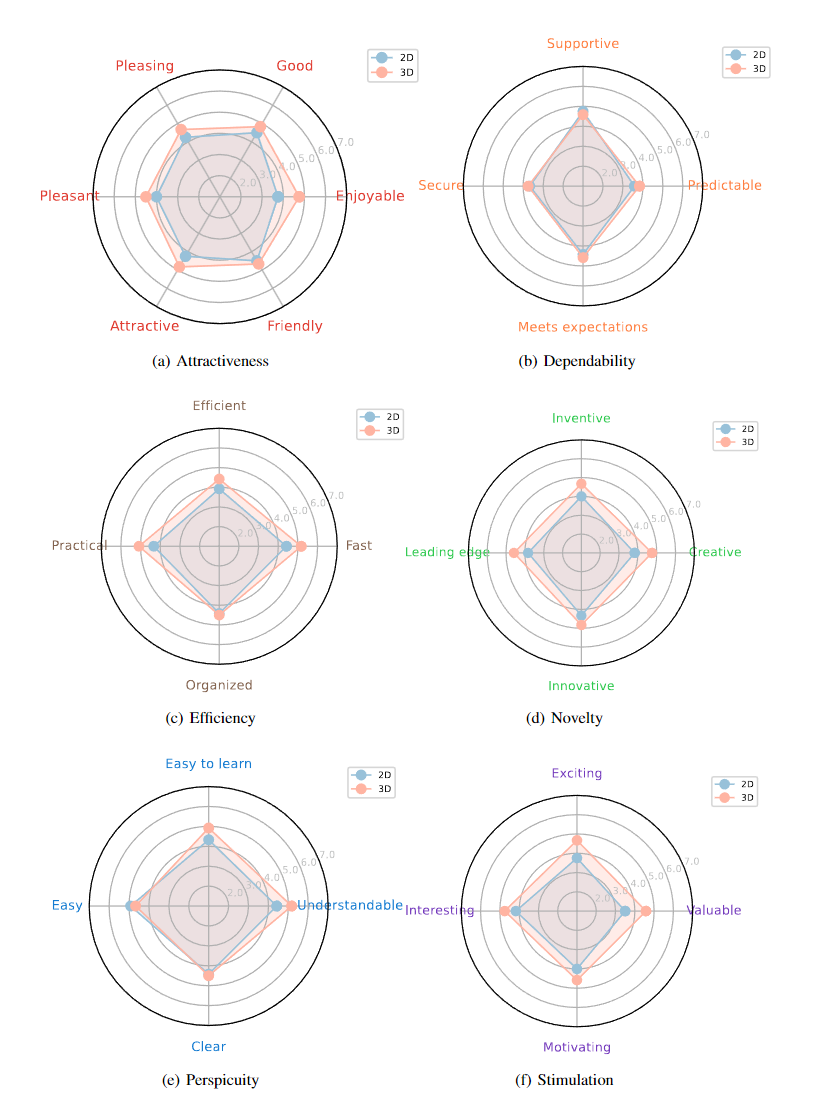}
    \caption{Each item of six scales of user experience.}    
\end{figure}

\begin{table}[ht]
  \centering
  \caption{Means and p-values for the three groups of scales.}
  \label{tab:p-value}
  \scalebox{0.85}{
  \begin{tabular}{ llll }
  \toprule[2pt]
  \makecell[c]{\textbf{Pragmatic and Hedonic Quality}} & \makecell[c]{\textbf{2D}} & \makecell[c]{\textbf{3D}} & \makecell[c]{\textbf{P-Value}}\\
  \midrule
  \makecell[c]{Attractiveness} & \makecell[c]{4.21} & \makecell[c]{4.71} & \makecell[c]{1.304E-03}\\
  \makecell[c]{Pragmatic Quality} & \makecell[c]{4.30} & \makecell[c]{4.60} & \makecell[c]{1.150E-02}\\
  \makecell[c]{Hedonic Quality} & \makecell[c]{3.93} & \makecell[c]{4.68} & \makecell[c]{5.056E-09}\\
  \bottomrule[2pt]
  \end{tabular}
  }
\end{table}

In the study mentioned in Schrepp \textit{et al.} \cite{schrepp2017construction}, the scales of the user experience questionnaire were categorised into two groups: pragmatic quality and hedonic quality. Pragmatic quality includes perspicuity, efficiency, and dependability, which are related to task performance. Hedonic quality includes stimulation and novelty, which are related to non-task-related aspects. Attractiveness, on the other hand, is considered a pure valence dimension. Table~\ref{tab:p-value} presents the average scores for these grouped scales for two different games, along with the corresponding p-values. The results indicate that the user preference for the 3D game was statistically significant, especially regarding hedonic quality.

\subsection{Comparing User Performance}

Table~\ref{tab:game scores} summarises the results, for each of the four user groups and for each of their two attempts on the game, separately. 
\begin{table}[ht]
  \centering
  \caption{Score means and standard deviations, for the four groups, for the first (I) and the second (II) attempts.}
  \label{tab:game scores}
  \scalebox{0.9}{
  \begin{tabular}{ lllll }
  \toprule[2pt]
\makecell[c]{\textbf{Groups}} & \makecell[c]{\textbf{mean (I)}} & \makecell[c]{\textbf{s.d. (I)}} &\makecell[c]{\textbf{mean (II)}} &\makecell[c]{\textbf{s.d (II)}}\\
\hline
\makecell[c]{2D-2D} &  \makecell[c]{10.50}& \makecell[c]{2.22} & \makecell[c]{17.33} & \makecell[c]{2.92}\\
\makecell[c]{2D-3D} &  \makecell[c]{9.83}& \makecell[c]{1.34} & \makecell[c]{14.67} & \makecell[c]{2.49}\\
\hline
\makecell[c]{3D-2D} &  \makecell[c]{9.33}& \makecell[c]{1.70} & \makecell[c]{14.83}& \makecell[c]{1.95}\\
\makecell[c]{3D-3D} &  \makecell[c]{9.67}& \makecell[c]{1.49} & \makecell[c]{16.00}& \makecell[c]{2.58}\\
\bottomrule[2pt]
  \end{tabular}
  }
\end{table}

To measure the effect of the choice between 2D and 3D, we compared scores over the two environments, on the first attempt only. These scores correspond to the first column of the table, and we compare the first two rows against the last two. The t-test returned a p-value of 0.43, indicating no statistical significance between the two environments, even though the 2D environment had a slightly higher mean score, 10.16 against 9.50. 

To measure the general learning effect, we compared the scores between the first and the second attempt, which correspond to the first and third columns of the table. The t-test returned a p-value of 2.48E-14, with means of 9.83 and 15.70 for the first and the second attempt, respectively, indicating a very strong learning effect. Notice that the design of the experiment is symmetric, and thus, the result was not affected by the order in which the 2D and 3D games were played. Moreover, by comparing the second column of the table with the fourth, we notice that in the second attempt the standard deviations increased within all groups, indicating that the learning effect was not uniform across all users. This is an observation that requires a larger-scale experiment to assess its significance. 

Finally, to specifically detect user familiarisation with the VR environment, as opposed to a general learning effect, we compared the second attempt scores between users who swapped environments between attempts (groups 2D-3D and 3D-2D) and those who did not (groups 2D-2D and 3D-3D). These scores correspond to the third row of the table, and we compare rows 1 and 4, against rows 2 and 3. The corresponding means were 14.75 and 16.66, and a t-test p-value of 0.043 indicates that the detrimental effect of swapping VR environments was small but statistically significant.




\section{Discussion}\label{sec:Discussion}

When examining the scales related to enjoyment and engagement, the scores show that users had a positive perception of both 2D and 3D games. The 3D games received higher ratings in terms of being enjoyable, good, pleasing, and attractive, indicating that they provided a more captivating and satisfying gaming experience. However, the 2D games also received respectable scores, suggesting that they still managed to offer a satisfactory level of enjoyment and engagement. These results indicate that while 3D games may have an edge in terms of overall enjoyment, 2D games can still provide an enjoyable gaming experience for users.

The scores related to learning and novelty reveal interesting insights about 2D and 3D games. The 3D games received higher ratings in terms of novelty and innovation, suggesting that they provided a fresh and groundbreaking gaming experience. This can be attributed to the immersive nature of 3D environments, which inherently offer a sense of novelty and exploration. Conversely, the 2D games were perceived as easier to use, indicating that they may have a lower learning curve and are more accessible to newcomers. These findings indicate that 3D games excel in terms of offering novel experiences, while 2D games provide a faster learning curve. Overall, the analysis suggests that the 3D game excelled in visual appeal, efficiency, stimulation, enjoyment, and novelty, while, on the other hand, the 2D game demonstrated strengths in interface clarity, perspicuity, and ease of learning. 

The quantitative analysis of the game scores did not find a statistically significant difference between 2D and 3D. However, there was a statistically significant learning effect, part at least of which should be explained by the familiarisation with the VR environment, rather than the learning of ASL. Moreover, there was an indication of different learning patterns among users, with the standard deviations increasing in the second attempt within all user groups. 

\subsection{Limitations}
It is important to acknowledge that both 2D and 3D games have their respective limitations. In the case of 2D games, one area that warrants future work is the enhancement of the visual experience. Advances in graphics technology could potentially narrow the gap between 2D and 3D games by allowing for more detailed and visually appealing 2D environments. Additionally, further exploration of gameplay mechanics and interactivity in the context of 2D games could lead to more engaging and immersive experiences for players. On the other hand, 3D games face a primary limitation related to hardware requirements. Future work and development efforts could focus on optimising game engines and graphics pipelines to improve performance on a wider range of devices. This would ensure that more players have access to and can enjoy 3D gaming experiences, regardless of their shareware capabilities. Additionally, it is important to note several limitations of our evaluation approach. One notable limitation is the relatively small sample size of 24 participants in the user study, which may have resulted in limited data and potential biases. To mitigate this limitation, future studies should consider increasing the sample size to obtain more robust and generalisable results.

\section{Conclusion and Future Work}\label{sec:Conclusion}
In conclusion, our study provides preliminary evidence of the potential of both 2D and 3D games in enhancing the user experience of learning ASL. The findings highlight the positive impact of 3D game environments on user engagement and overall experience, as evidenced by their higher ratings in attractiveness, usability, and efficiency, compared to 2D games. However, there is room for improvement in ensuring the dependability and clarity of 3D game environments. These results contribute to our understanding of the benefits of incorporating game-based approaches, particularly 3D environments, into ASL learning. 

Future research can build upon these findings by delving deeper into the specific elements and design features that contribute to the positive user experience in both 2D and 3D games. Additionally, exploring strategies to enhance the dependability and clarity of 3D game environments can further optimise learning outcomes in ASL education. It would also be beneficial to increase the number of participants in future user studies to strengthen the generalisability of the findings.


\bibliographystyle{unsrt}  
\bibliography{references}

\end{document}